\documentclass{IOS-Book-Article}     %[seceqn,secfloat,secthm]

\usepackage{amssymb}
\usepackage{graphicx}
\usepackage{siunitx}
\usepackage{mathtools}
\usepackage{wrapfig}
\usepackage{lipsum}
\usepackage{cite}
\usepackage{marvosym}
\usepackage{soul}
\usepackage{xcolor}
\usepackage{float}

\begin{document}
   \begin{frontmatter}          % The preamble begins here.
      %
      %\pretitle{}
      \title{Dense Fiber Modeling for 3D-Polarized Light Imaging Simulations}
      \runningtitle{Dense Fiber Modeling for 3D-PLI Simulations}
      %\subtitle{}
      
      % Two or more authors:
      \author[A]{\fnms{Felix} \snm{Matuschke}},
      \author[B]{\fnms{K\'{e}vin} \snm{Ginsburger}},
      \author[B]{\fnms{Cyril} \snm{Poupon}},
      \author[A]{\fnms{Katrin} \snm{Amunts}},
      \author[A]{\fnms{Markus} \snm{Axer}}
      \runningauthor{F. Matuschke et. al.}
      \address[A]{Institute of Neurosience and Medicine (INM-1),\\ Forschungszentrum J\"{u}lich, 52428 J\"{u}lich, Germany}
      \address[B]{CEA DRF/ISVFJ/Neurospin/UNIRS, Gif-sur-Yvette, France}
      \begin{abstract}
         3D-Polarized Light Imaging (3D-PLI) is a neuroimaging technique used to study the structural connectivity of the human brain at the meso- and microscale. In 3D-PLI, the complex nerve fiber architecture of the brain is characterized by 3D orientation vector fields that are derived from birefringence measurements of unstained histological brain sections by means of an effective physical model.
         
         To optimize the physical model and to better understand the underlying microstructure, numerical simulations are essential tools to optimize the used physical model and to understand the underlying microstructure in detail. The simulations rely on predefined configurations of nerve fiber models (e.g. crossing, kissing, or complex intermingling), their physical properties, as well as the physical properties of the employed optical system to model the entire 3D-PLI measurement. By comparing the simulation and experimental results, possible misinterpretations in the fiber reconstruction process of 3D-PLI can be identified. Here, we focus on fiber modeling with a specific emphasize on the generation of dense fiber distributions as found in the human brain's white matter. A new algorithm will be introduced that allows to control possible intersections of computationally grown fiber structures.
      \end{abstract}
      
      
   \end{frontmatter}
   
   %%%%%%%%%%% The article body starts:
   
   \section{Introduction}
   3D-Polarized Light Imaging (3D-PLI) is a unique microscopy technique to study the nerve fiber architecture  of human brains as well as brains of non-human primates and rodents at micrometer scale \cite{Axer2011a, Axer2011b, Larsen2007, Zilles2016}. In 3D-PLI, predefined polarized light is passed through unstained $\SI{60}{\micro\meter}$ thick histological brain sections and detected by a camera. Due to the optical property by birefringence as a property of nervous tissue, in particular to myelinated axons (in the following referred to as "nerve fibers"), the polarization state of light alters during the passage of light through the section. The total change of the polarization state of light is measured in 3D-PLI and enables to derive a 3D orientation vector per measured voxel (defined by image pixel size and section thickness) from an effective birefringence model \cite{Goethling}. Such orientation vector is referred to as fiber orientation. The way the polarization state of light is changed depends on a complex interplay of various tissue properties, including fiber density, fiber course or orientations, fiber thickness and degree of myelination. The currently used physical model does not allow to separate these parameters unambiguously, which makes the determination of fiber orientations in some cases prone to misinterpretation.
   
   To better understand how the derived orientation vectors are related to the underlying fiber architecture and to improve the accuracy of the reconstructed fiber orientations, analytical methods and numerical simulations were developed \cite{Dohmen2015,Menzel2015, Menzel2016}. Depending on the studied physical effect, two general types of simulations were pursued to model the interaction of light with brain tissue: The birefringence of the nerve fibers was simulated by means of the Jones \cite{Jones1941, ClarkJones1942} or the M\"uller-Stokes matrix calculus \cite{memorandum} implemented in an in-house developed framework called \emph{simPLI} \cite{Dohmen2015, Menzel2015}. To investigate optical effects, such as scattering and interference, a massively parallel 3D Maxwell solver is used that is based on a \emph{finite-difference time-domain} (FDTD) algorithm \cite{Taflove,Menzel2016,1806.07157}. While \emph{simPLI} runs efficiently on the cluster supercomputer \emph{JURECA} \cite{jureca}, the 3D Maxwell solver simulations utilized the advantages of  the high-scaling supercomputer \emph{JUQUEEN} \cite{juqueen} at the J\"ulich Supercomputing Centre (JSC) at the Forschungszentrum J\"ulich, Germany.
   
   The uniting elements for both types of simulation were well-defined input fiber models, which were typically realized as parallel symmetrical tubes \cite{Dohmen2015}. However, these models were susceptible to symmetry effects leading to interference patterns in the resulting simulations as described in wave optics \cite{Menzel2016}. Clearly, there was need for new tools (i) to render controlled and defined generation of more complex fiber models possible and (ii) to make the fiber models suitable for 3D-PLI simulations. The modelling of hundreds to thousands of non-intersecting fiber structures within a given voxel is a most computation-intensive process, which requires specific attention.
   
   In the following, the creation of dense fiber models as representatives of white matter fiber bundles. In the human brain, fiber bundles can be composed of a few tens of fibers (e.g., resulting in the myeloarchitectonic pattern in the cerebral cortex \cite{Nieuwenhuys2012}), but also of hundreds to millions of fibers forming large fiber bundles (e.g., the corpus callosum \cite{Luders2010}). Fiber bundles may cross in various constellations (e.g., stacked, interwoven, under different crossing angles) or they might merely touch each other ("kissing fibers"). In order to analyze the effects caused by different fiber constellations, fiber modeling has to be valid and controlled. In addition, the resulting models have to be suited to be analyzed with the same tools or workflows, resp., that were specifically developed for experimental 3D-PLI data \cite{Amunts2014}.
   
   Especially in the case of high-density nerve fiber bundles, it is important to prevent interpenetrating fibers during their generation as they inevitably take effect on the simulated results. This is not only true for 3D-PLI simulation approaches \cite{Menzel2016}, but also for water diffusion modeling in Magnetic Resonance Imaging (dMRI). Here, for example, similar types of fiber models have been used to study the microstructural impact on the (un-)restricted water diffusion in the brain \cite{Ginsburger2018, Estournet2017}.
   
   %------------------------------------------------
   
   \section{Generation of nerve fiber models}
   
   \begin{figure}[!b]
      \centering
      \includegraphics[width = \textwidth]{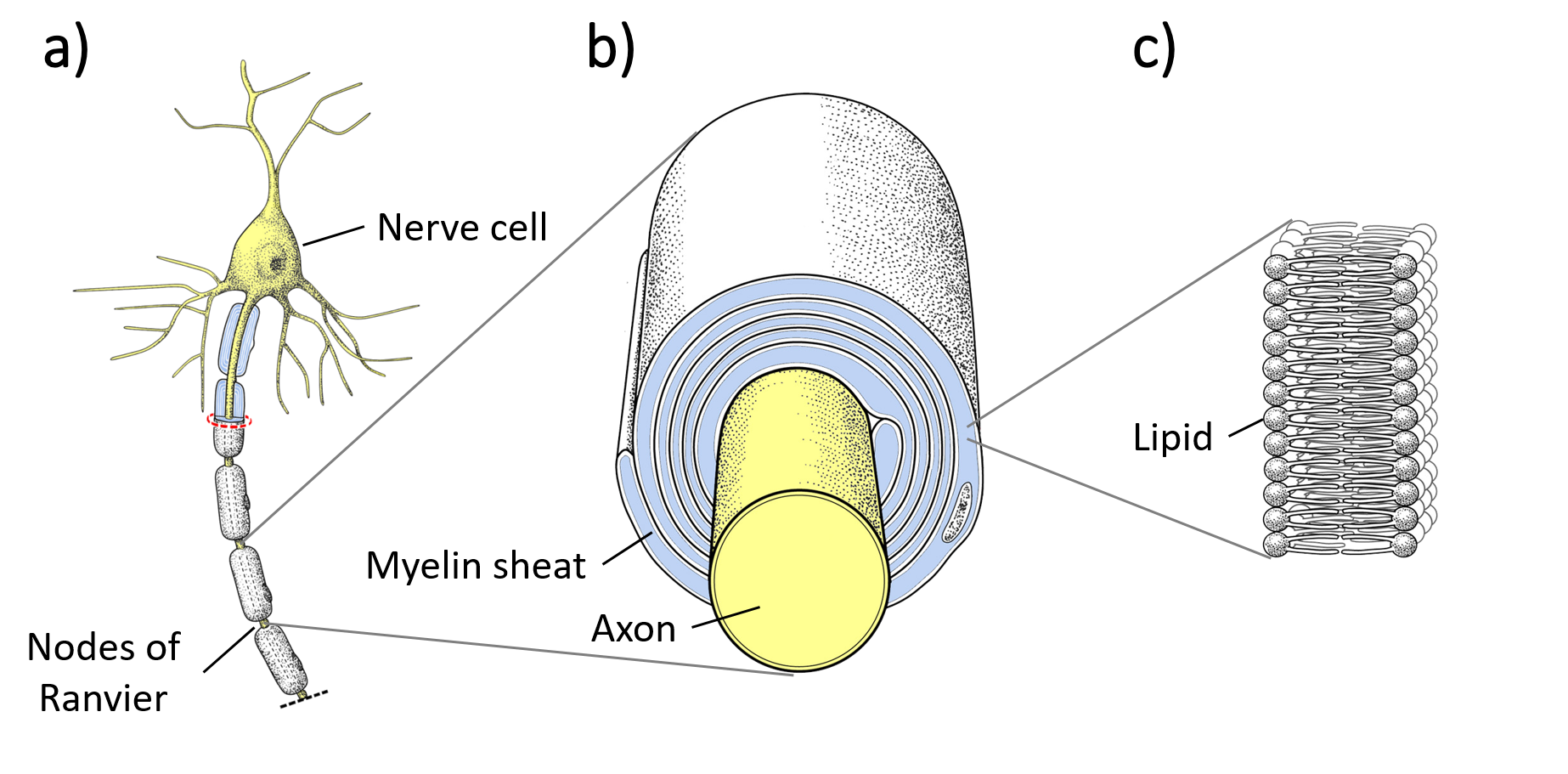}
      \caption{
         a) Schematic structure of a nerve cell with myelinized axon. b) Cross section of a nerve fiber with spirally covered myelin. c) The myelin consists of double lipid membranes. (Copyright: INM-1, FZJ, Germany)}
      \label{fig:NerveFiberStructure}
   \end{figure}
   
   Single myelinated nerve fibers consists of an axon and a myelin sheath (see fig. \ref{fig:NerveFiberStructure}). The myelin sheath and the axon can vary in the thickness from fiber to fiber. The myelin is divided into different sections, separated by the nodes of Ranvier. Such structure is the basis for saltatory signal transduction. The myelin itself consists mainly of a combination of a lipid double membrane and proteins. The human axon diameter $d_\text{axon}$ range from about  $\text{\SIrange{0.2}{5}{\micro\meter}}$, the myelin thickness $t_\text{myelin}$ from about $\text{\SIrange{0.05}{0.35}{\micro\meter}}$ \cite{FitzGibbon2013}. 
   
   In order to model a single nerve fiber three-dimensionally, different strategies can be pursued. One strategy is to describe the nerve fiber as a parameterized three-dimensional curve $f: \mathbb R \rightarrow \mathbb R^3$, e. g. $f(t) = (t,0,0)$ for a horizontal fiber. A radius $r(t)$ provides the fiber with a cylindrical shape corresponding to the envelope of the nerve fiber with myelin. This parametric representation allows to describe all kinds of cylindrical shapes. However, designing bundles of non-colliding nerve fibers remains challenging. While a straight fiber bundle is simple to design, it is not a trivial task to ensure that adjacent fibers do not collide for curved geometries.
   
   This problem is avoided using an algorithm that places one fiber after the other into a volume and checks each time whether the new fiber collides with any already existing fiber. If a collision is detected, the lastly placed fiber can be re-positioned. However, this type of algorithm has the decisive disadvantage that it becomes more and more difficult for a given volume to fill it with new fibers. It becomes even more problematic, if two nerve fiber bundles are supposed to intersect each other, as it occurs, for example, in the optic chiasm \cite{Dohmen2015}.
   
   For these reasons, a different approach is proposed. A nerve fiber is divided into equal sections along its defined function. This corresponds to a chain composed of linked cylindrical segments (see fig. \ref{fig:algorithem} a) and b)). Now it is possible to move individual colliding chain links to new positions to generate non-colliding fibers. The algorithm to resolve these collisions is presented in section \ref{chap::FiberModelling}.

\begin{figure}[!t]
      \centering
      \includegraphics[width = 0.9\textwidth]{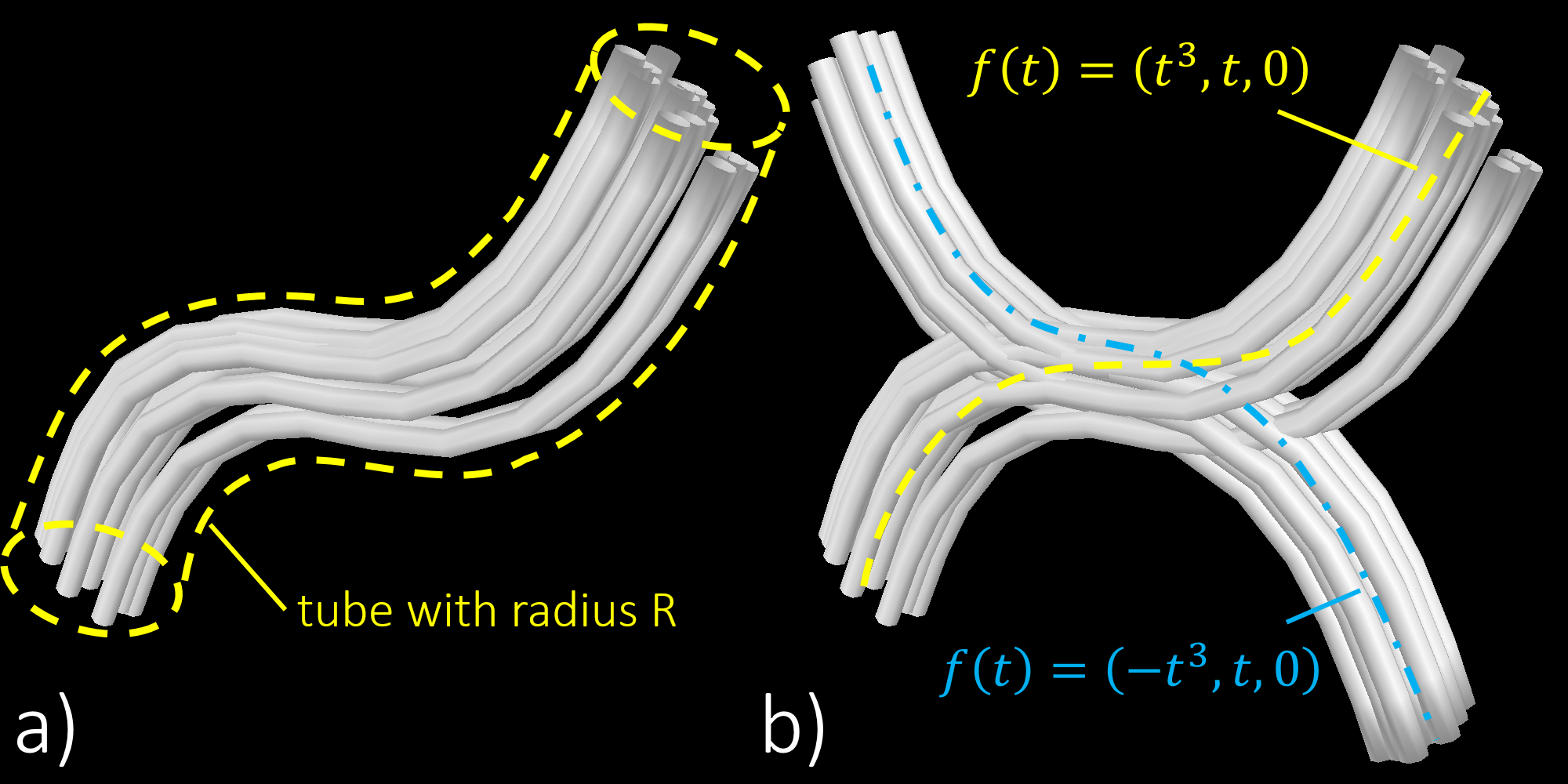}
      \caption{a) A single nerve fiber bundle is created along the function $f(t) = (t^3,t,0)$. The individual nerve fibers are placed randomly within the global bundle radius $R$. b) A second, mirrored nerve fiber bundle is added to the first bundle. The individual fibers are colliding with each other.}
      \label{fig:BundleExample}
   \end{figure}

   \paragraph{Example:} Let's assume, the basic form of the fiber bundle is given by the function $f (t) = (t^3, t, 0)$ (fig. \ref{fig:BundleExample} a)). This bundle has a radius $R$ and is filled with single fibers of equal radius $r$. In this example, the number of fibers in the bundle was set to $N = 10$. Each fiber is positioned randomly within its radius $R$ around its center $(y,z)$ position. At this step, individual fibers may collide within the bundle. More complex structures (e.g., intersecting bundles, cf. \ref{fig:BundleExample} b)) can be assembled from small elementary bundles such as the one shown in figure \ref{fig:BundleExample}a).
   
   %------------------------------------------------
   
   \section{Disentangling colliding fibers} \label{chap::FiberModelling}
   
   The goal is to ensure that no single fiber collides with each other for any given fiber configuration. The algorithm is based on the approaches proposed by the work of Chapelle et.al \cite{Chapelle2015} and Altendorn \& Jeulin \cite{Altendorf2011}. Both used collision detection of geometrical shapes to create a 3D fiber model. Chapelle et.al use cone shapes to approximate fibers. On the other side Altendorn \& Jeulin use sphere objects. Both use different kinds of forces to solve the collision state and keep the fibers in shape. The algorithm presented here uses concepts from their research as well as additional elements to create collision-free models for 3D-PLI.
   
\begin{figure}[!t]
      \centering
      \includegraphics[width=0.5\textwidth]{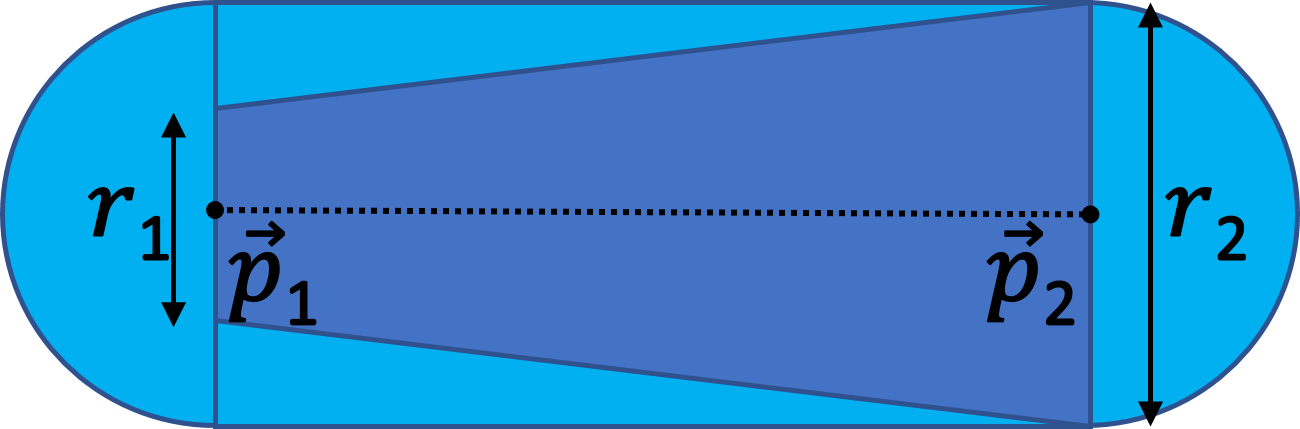}
      \caption{Each fiber segment is defined by two points $\vec{p}_1$ and $\vec{p}_2$ and their corresponding radii $r_1$ and $r_2$. The corresponding shape can be represented by a cone (dark). In order to simplify the collision calculations, the segment is enclosed in a capsule (bright).
      }
      \label{fig:shape}
   \end{figure}

\begin{figure}[p]
      \centering
      \includegraphics[width = \textwidth]{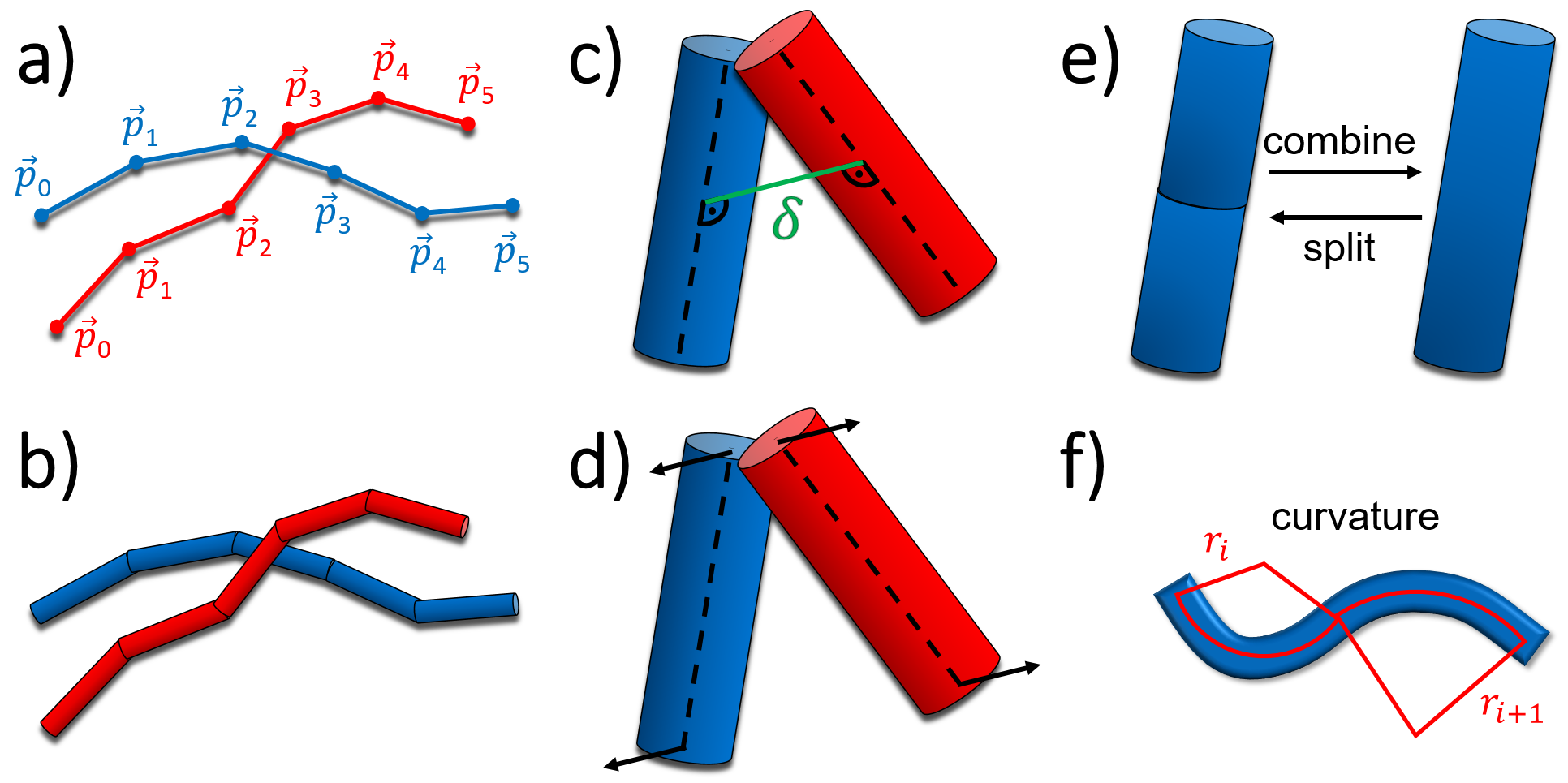}
      \caption{Different steps of the collision solver algorithm. a) Fibers are represented as a list of coordinates $\vec{p}_i$. b) Segments between two points $\vec{p}_i$ and $\vec{p}_{i+1}$  and their radii $r_i$  and $r_{i+1}$ build a 3D chain of cones. c) To detect a collision the smallest distance $\delta$ between two capsule objects is calculated. d) When a collision is detected, both objects are moved apart. e) The number of objects can be adjusted. It can divide an elongated segment into two parts or combine two shortened segments into one. f) Measurement of the curvature via the circumradius. If the circumradius is too small, the middle point is shifted towards the circle center so that the radius grows and the curvature becomes smoother.
      }
      \label{fig:algorithem}
      
      \vspace*{\floatsep}
      
      \includegraphics[width=1\textwidth]{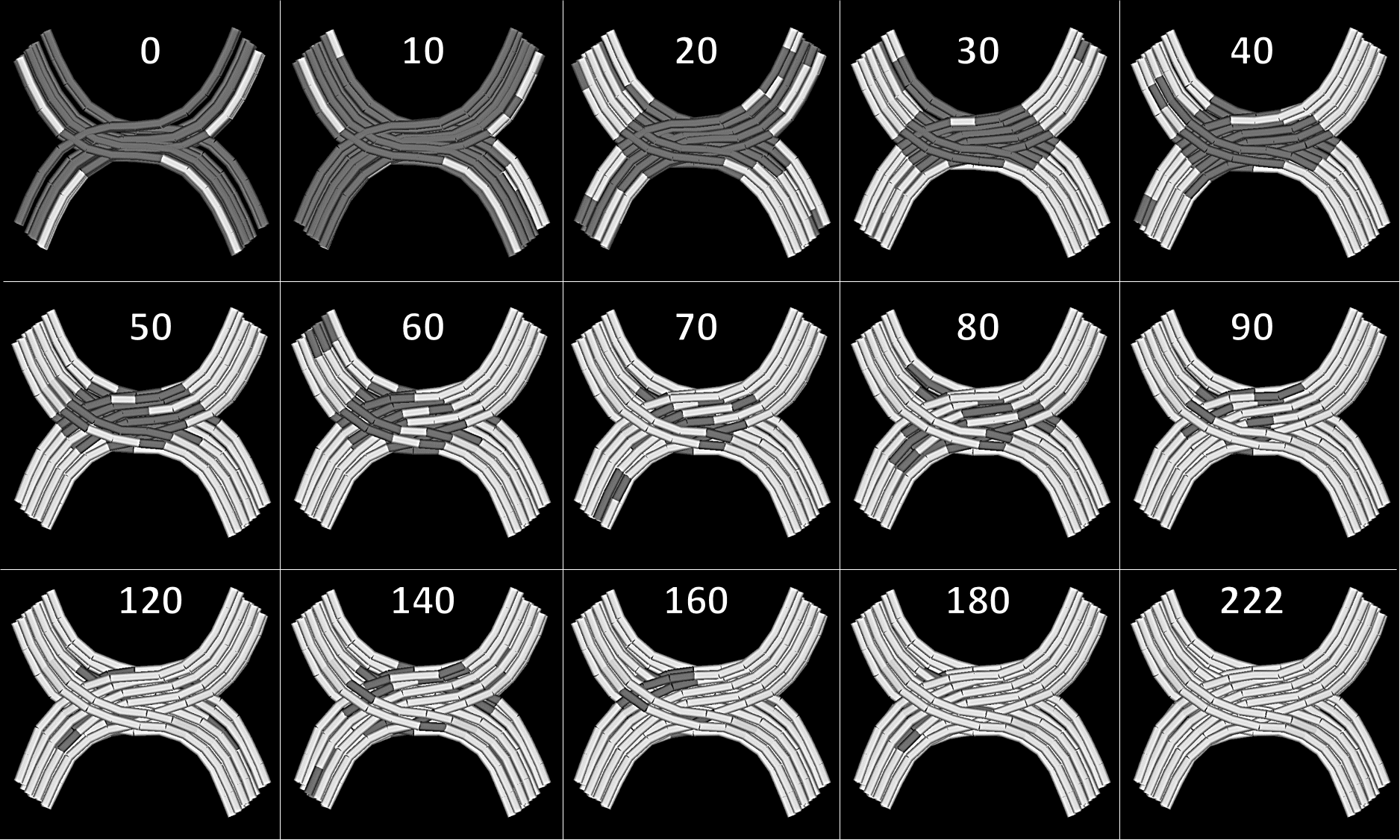}
      \caption{Solution of the example from fig. \ref{fig:BundleExample} by the algorithm presented here. The two bundles are deformed in a total of 222 steps in such a way that no more collision takes place. Collisions are represented here as dark grey segments.}
      \label{fig:colision}
   \end{figure}

   Fibers are represented as an array of tuples $(x_i,y_i,z_i,r_i)$ containing the three spatial coordinates $\vec{p}_i = (x_i,y_i,z_i)$ and the fiber radius $r_i$ at each point $\vec{p}_i$ (see fig. \ref{fig:algorithem} a)). As the radius may change along the fiber, e. g., to build nodes of Ranvier, each chain link $(\vec{p}_i, \vec{p}_{i+1})$ corresponds to a conical object (see fig. \ref{fig:shape}). 
   
   \newpage
   
   \subsection{Collision Detection}
   The capsule representation (see fig. \ref{fig:shape}) is used instead of cone shapes to speed up collision detection between two fiber segments. This means that only the larger radius of the object is used:
   \begin{equation}
   r^{\text{caps}} = \max(r_1^{\text{cone}}, r_2^{\text{cone}}) .
   \end{equation}
   Therefore the object is now represented as a chain of capsules. This is a reasonable choice for nerve fibers as the radius is relatively small compared to adjacent segments. To check whether two capsules collide, the shortest Euclidean distance between two geometrical line segments $\left( \vec{p}_i,\vec{p}_{i+1} \right)$  and $\left( \vec{p}_j,\vec{p}_{j+1} \right)$ is calculated (see fig. \ref{fig:algorithem} c)). If the smallest distance $\delta$ between two segments $i$ and $j$ is smaller than the sum of the radii of each capsule $\delta < r_i^{\text   {caps}} + r_j^{\text{caps}}$, both objects are marked as colliding and placed in a list of colliding objects to start the separation process in the next step.
   
   \subsection{Separation Process}
   To separate two objects $i$ and $j$ from each other, each fiber point is assigned a velocity $\vec{v}_i$ (see fig. \ref{fig:algorithem} d)). At the beginning of each iteration the velocity of every point is set to zero. If two objects collide, both points of each object are increased by a speed of $|\vec{v}| = \SI{0.01}{\micro\meter/step}$ parallel to the minimal distance vector $\vec{\delta}$ away from the colliding object, therefor $\vec{v}_i$ points from $j$ to $i$ along $\vec{\delta}$ and vice versa (see fig \ref{fig:algorithem} d)). Since one object can collide with multiple objects, all velocities are summed up before moving each point. Each velocity will be restricted to a maximum of $\SI{0.1}{\micro\meter/step}$ so that the objects don't move too far and the fibers get entangled. This value was chosen because it is about one order of magnitude smaller than the objects to be moved. The first and last point of each fiber is treated individually. To ensure that the fibers don't shrink or enlarge, the endpoints are only moved perpendicular to their segment line.
   
   \subsection{Shape Control}
   Since the points of a fiber can move individually, it is reasonable to control the length between two neighboring points (see fig. \ref{fig:algorithem} e)).  The algorithm restricts the segment length in a range of $\left[l_\text{min},l_\text{max} = 2 \cdot l_\text{min} \right]$ and a mean value of $\bar{l} = \frac{3}{2} \cdot l_\text{min}$. If a segment is smaller or longer than the limits, it is combined with its neighbors or divided into two new segments. The targeting length of each segment can be controlled individually so that structures with rapidly changing radii do not disappear, e.g. such as found in nodes of Ranvier. To control the curvature of the individual fibers, the circumscribed circle and its radius $r$ are calculated from each point via its two neighbors:
   
   \begin{equation}\label{eq:2}
   \begin{split}
   a =& || \vec{p}_{i} - \vec{p}_{i-1} || \\
   b =& || \vec{p}_{i+1} - \vec{p}_{i} || \\
   c =& || \vec{p}_{i-1} - \vec{p}_{i+1} ||
   \end{split}
   \hspace{2em}
   \begin{split}
   \begin{aligned}
   s =& (a+b+c)/2 \\
   f =& \sqrt{s(s-a)(s-b)(s-c)} \\
   r =& a b c/(4f)
   \end{aligned}
   \end{split}
   \end{equation}
   
   This allows an object length independent measurement of its curvature (see fig. \ref{fig:algorithem} f)). If the circumscribed circle is smaller than a user defined minimal radius $r_\text{min}$, the points will be flattened, so that the curvature reduces. This is an efficient way to guarantee the smoothness of the geometry and of the fiber. By defining a minimal radius for the curvature each fiber point can be smoothed if necessary, so that e.g. strong direction changes like in the gray matter would be possible. However, for a densely packed fiber bundle it is feasible to use a global value.
   
   \subsection{Implementation} \label{sec:implementation}
   
   \begin{figure}[!b]
      
      \includegraphics[page=1,width=.5\textwidth]{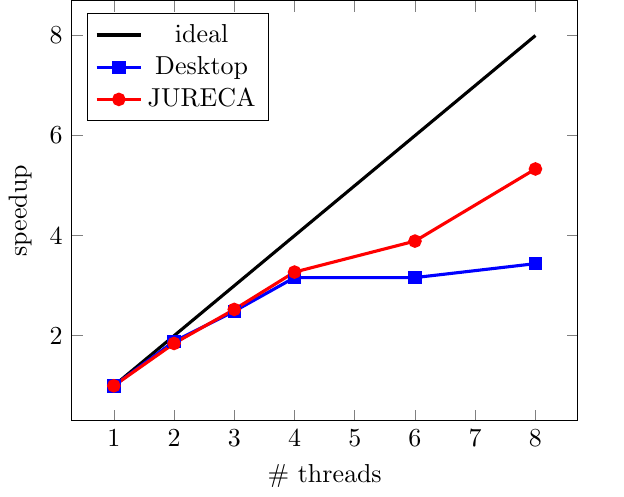}
      \caption{Speedup curve of the algorithm with the JURECA (Intel Xeon E5-2680) and a Desktop (Intel Core i7-3770) system. The Desktop system shows a plateau after 4 threads. The JURECA system shows a slower growth after 4 and a jump with 8 cores.
      }
      \label{fig::speedup}
   \end{figure}
   
\begin{figure}[!t]
      \centering
      \fbox{
         \includegraphics[page=1,width=.75\textwidth]{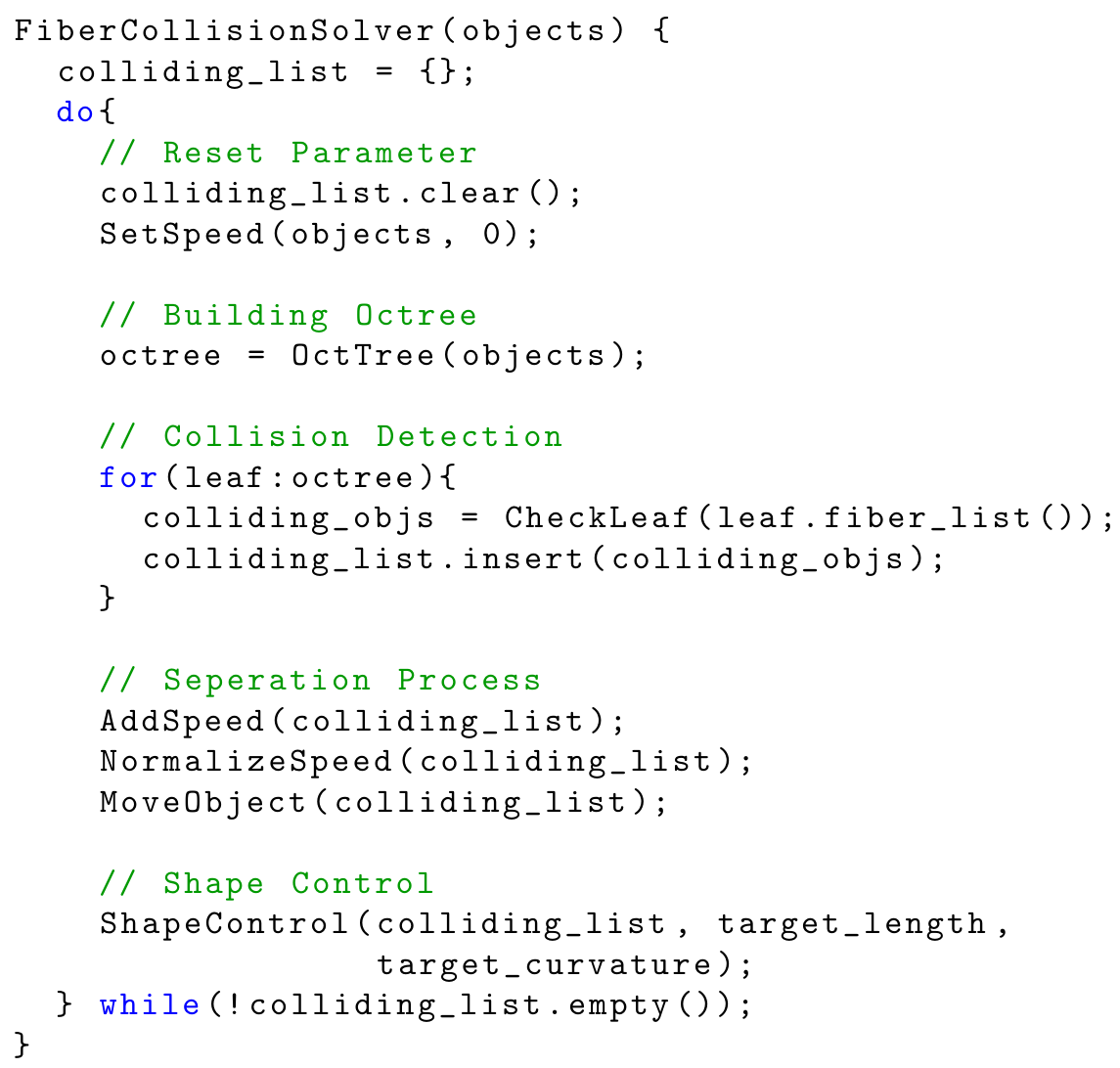}
      }
      \caption{Pseudocode of the main algorithm: The function \texttt{FiberCollisionSolver} will loop the followings four steps, which are run in parallel, until no collision are detected anymore: 1. build an \texttt{OctTree} from all objects, 2. \texttt{Collision Detection}, 3. \texttt{Seperation Process} and 4. \texttt{Shape Control}.}
      \label{alg::pseudocode}
   \end{figure}

   The collision detection, separation process and angle control are repeated until no more collisions are detected. \emph{OpenGL} is used to visualize the fiber configurations. An example is shown in figure \ref{fig:colision}. The current implementation is in \emph{C\texttt{++}}. To accelerate collision detection, an \emph{octree} with a complexity of $O(N \log N)$ is used. Furthermore, to parallelize the calculations \emph{OpenMP} is used for all thread safe operations. Currently the code is optimized for 8 CPU threads. The pseudocode of the algorithm is shown in fig. \ref{alg::pseudocode}.
   
   Depending on the architecture, measurement of speedup with multiple threads shows that the algorithm has a speedup about $3.3$ for 4 threads and $5.3$ for 8 threads (see fig. \ref{fig::speedup}). A desktop system (Intel Core i7-3770) and JURECA (Intel Xeon E5-2680) were used for the tests. An additional increase of the number of threads does not significantly increase the performance, since the current implementation of the \emph{octree} is optimized for 8 CPU threads. Separation process and shape control are processed independently in parallel.
   
   \subsection{Exemplary Tissue Models}
\begin{figure}[!t]
      \centering
      \includegraphics[width = \textwidth]{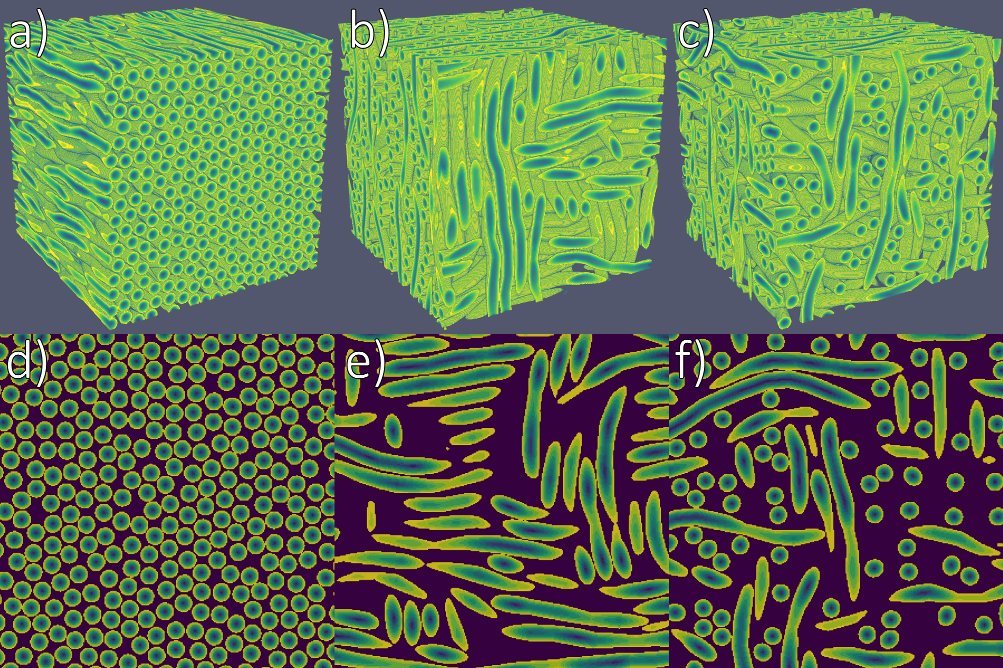}
      \caption{Three different crossing fibers in a volume of $\SI{30x30x30}{\micro\meter}$. Top: a) unidirectional fiber block $(\phi_a = \SI{65}{\percent})$, b) two perpendicular crossing fiber populations $(\phi_b = \SI{52}{\percent})$, c) three perpendicular crossing fiber populations $(\phi_c = \SI{45}{\percent})$. Bottom: corresponding slices through the midsection. 
         (Visualization by \emph{ParaView} \cite{ParaView}.)
      }
      \label{fig:threecubes}
   \end{figure}

   We have tested the algorithm in different examples of dense fiber bundle of crossing fibers with one $(a)$, two $(b)$ and three $(c)$ fiber populations (Figure \ref{fig:threecubes}). A volume of $\sqrt{3^3} \cdot (\SI{30}{\micro\meter})^3$  is used with a total number of $\text{N}_{\text{fibers}} = \SI{1000}{}$ uniform randomly distributed fibers, which statistically collide. The final volume is reduced to $(\SI{30}{\micro\meter})^3$ to ensure that there are no boundary effects. The mean segment length is set to $\bar{l}=\SI{2}{\micro\meter}$. To create a tortuosity, at the beginning all points are displaced randomly with a uniform distribution with a maximum displacement of $\SI{0.5}{\micro\meter}$ in each direction. The radius of each fiber is constant with $r = \SI{0.8}{\micro\meter}$. Note that due to the image resolution in the cutting plane the fibers can apparently touch. 
   
   The resulting volume fractions $\phi$ are $\phi_{(a)} = \SI{65}{\percent}$, $\phi_{(b)} = \SI{52}{\percent}$ and $\varphi_{(c)} = \SI{45}{\percent}$. The mathematical maximum for cylinders are $\phi_{(a)}^\text{max} = \frac{\pi}{2 \sqrt{3}} \approx \SI{91}{\percent}$, $\phi_{(b)}^\text{max} = \frac{\pi}{4} \approx \SI{79}{\percent}$  and $\phi_{(c)}^\text{max} = \frac{3\pi}{16} \approx \SI{59}{\percent}$. The computational time for these volumes on a 8-cores system (Desktop or Jureca) is in the order of minutes.
   
\begin{figure}[t]
      \centering
      \includegraphics[width = \textwidth]{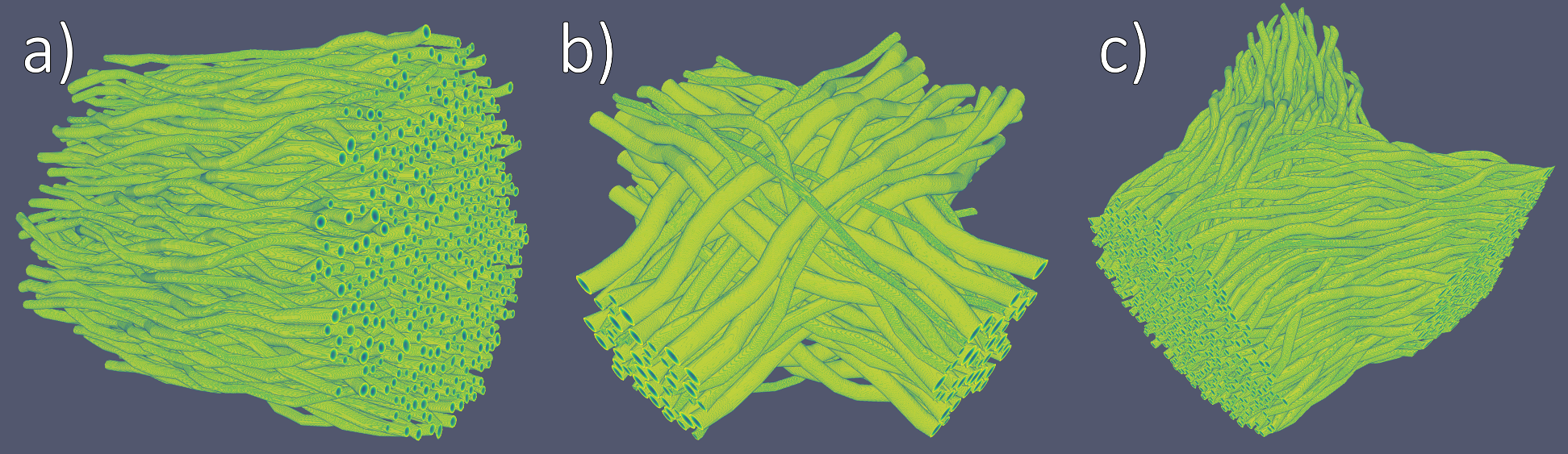}
      \caption{Exemplary fiber populations. a) cylindrical fiber bundle, b) crossing fiber bundle, c) Y-shape fiber bundle. All configurations are biologically meaningful, and can be found in many brain regions. (Visualization by \emph{ParaView} \cite{ParaView}.)}
      
      \label{fig:fourexamples}
   \end{figure}

   Figure \ref{fig:fourexamples} shows three examples of four different types of fiber geometries. It is possible to generate any kind of structure with variation of fiber radius, segment length, angle dispersion, etc.
   
%   \vspace{-2em}
   
   %------------------------------------------------
   
   \section{Using fiber models for 3D-PLI simuations}
   
   \subsection{The physics behind 3D-PLI}
\begin{figure}[!t]
      \centering
      \includegraphics[width = 1\textwidth]{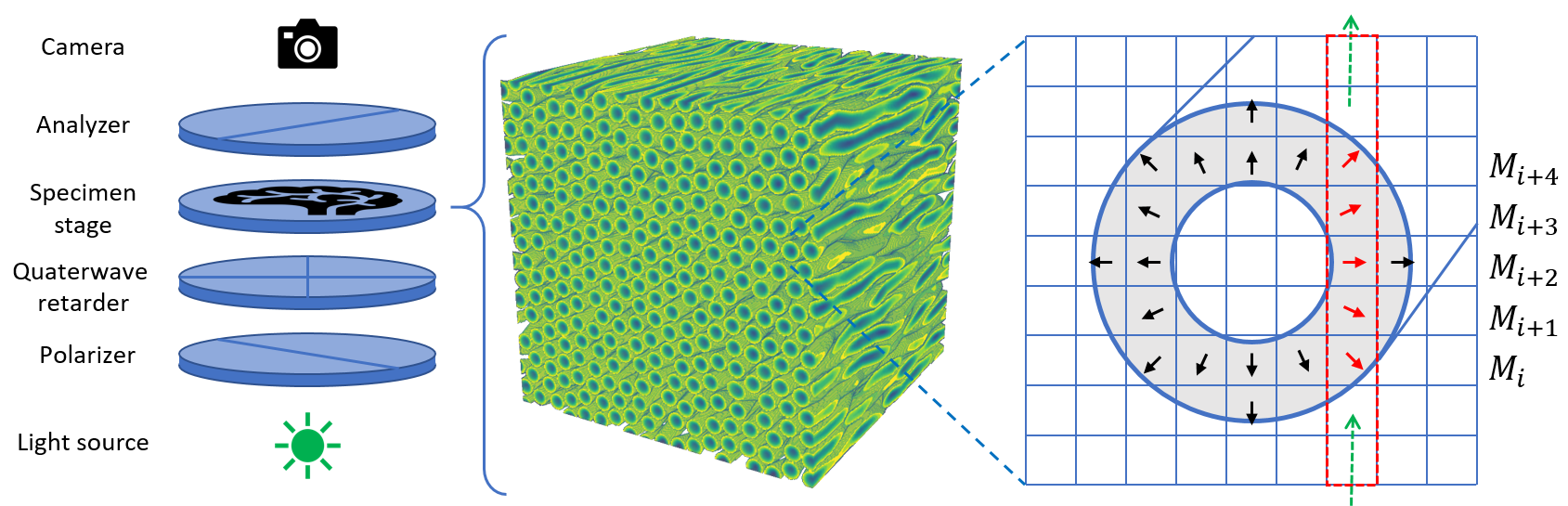}
      \caption{Experimental and simulative setup. The 3D-PLI setup consist of a light source, a polarizer, a quarter wave retarder, the specimen stage, a analyser and a camera. To simulate this setup, the Jones-Calculus is used in \emph{simPLI}. For this purpose tissue properties, such as birefringence and direction, are assigned to discrete voxels.}
      \label{fig:PLI}
   \end{figure}

   3D-PLI uses the birefringent properties of the nerve fibers to visualize their direction. A formalin-fixed brain is cryo-sectioned into $\text{\SIrange{60}{70}{\micro\meter}}$ thin sections and placed section-wise between two polarizers and a quarter-wave retarder (see fig. \ref{fig:PLI}). An LED panel ($\lambda = \SI{525}{\nano\meter}$) serves as a light source. For the measurement, the polarizer and quarter-wave retarder are simultaneously rotated in steps of \SI{10}{\degree} from $\text{\SIrange{0}{170}{\degree}}$. For each rotation angle $\rho$ an image $I(x,y,\rho)$ is acquired with a CCD-camera. In each pixel $I(x,y,\rho)$ shows a sinusoidal behaviour, which can be physically described as a uniaxial birefringent crystal with its principle optic axis aligned along the fiber axis \cite{Axer2011b}: 
   \begin{equation}
   I(\rho, \varphi, \alpha, t) = I_T \left(1+\sin\left(2(\rho-\varphi)\right) \sin\left(\frac{\pi}{2}t \cos^2(\alpha)\right)\right) \label{eq:PLI}
   \end{equation}
   with the intensity $I$, the rotation angle $\rho$, the fiber orientation $\varphi$, the inclination angle of the fiber $\alpha$, the relative fiber density $t_{\text{rel}}$ and the transmitted light intensity $I_T$. The relative fiber density is given by $t_{\text{rel}} = 4 t_s \Delta n / \lambda$ with $t_s$ as thickness, $\Delta n$ as birefringence and $\lambda$ as wavelength of the light. From equation (\ref{eq:PLI}) the three modality maps are processed for each pixel $(x,y)$: The \emph{transmittance map} $I_T(x,y)$, the \emph{direction map} $\varphi(x,y)$ and the \emph{retardation map} $ret(x,y) \coloneqq \sin(\frac{\pi}{2}t \cos^2(\alpha))$ via a Fourier transformation.
   
   \subsection{Simulation of 3D-PLI measurements}
   To generate a 3D-PLI simulation the optical axis along the path of the light has to be calculated. This is possible with the geometry of the generated fiber models. Two models have been described by \cite{Menzel2015}, the microscopic and macroscopic model. The microscopic model describes the birefringence caused by myelin as radially arranged layers surrounding the axon. The macroscopic model describes the axis along the nerve fiber. Since the used simulation uses a rather small grid size, the microscopic model is essential (see fig. \ref{fig:PLI}). 
      
   \begin{figure}[!t]
      \centering
      \includegraphics[width=\textwidth]{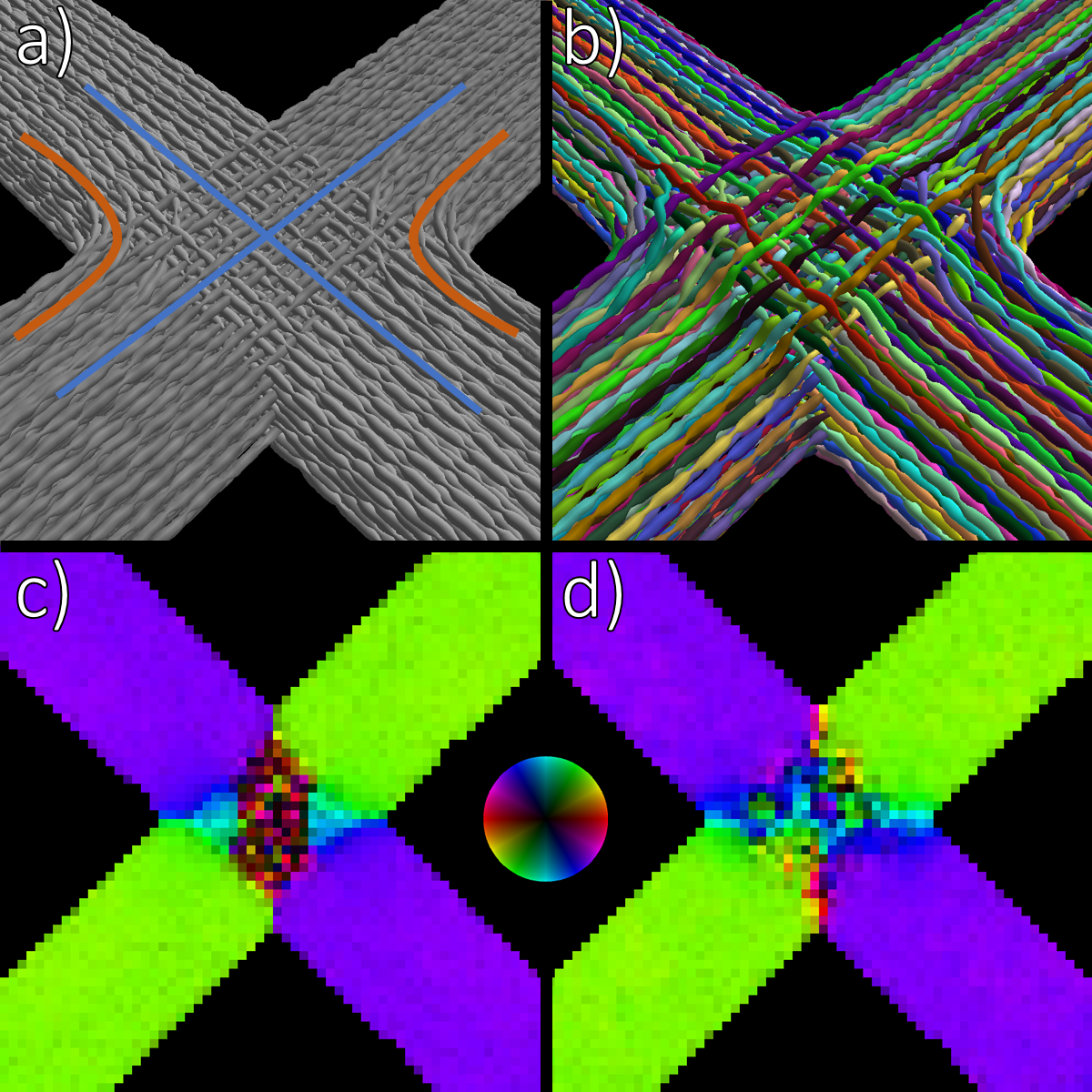}
      \caption{Example of the 3D-PLI simulation process: 
         a) Creation of initial crossing fibers (blue). This example is inspired by the optic chiasm. In addition to the crossing fiber bundles, a small proportion of the fibers also retain their sides (orange).
         b) Result of the colliding solving algorithm. Since the fibers have to be shifted in the intersection, the volume increases significantly in the middle.
         c) FOM of initial fiber constellation. The orientation in the center is distorted by unrealistic geometries.
         d) FOM of the collision-free volume. Compared to the previously FOM, the crossing area shows smoother transitions between the fiber bundles.
      }		
      \label{fig:sim_example}
   \end{figure}
   
   The in house developed tool \emph{simPLI} is used to simulate the 3D-PLI data from the modeled nerve fibers \cite{Dohmen2015, Menzel2015}. It uses linear optics to calculate the polarized state of the light through the birefringent medium. The simulated signal is calculated via the M\"uller calculus \cite{memorandum}. The essential equation to calculate the change of the electromagnetic field $\vec{E}$ through the 3D-PLI setup is:
   \begin{equation}
   \vec{E} = P_y \cdot (M_N \cdot M_{N-1} \cdots M_{1})  \cdot M_{\lambda/4} \cdot P_x \cdot \vec{E}_0
   \end{equation}
   
   $\vec{E}_0$ is the initial electromagnetic field. $P_x$ and $P_y$ represent the polarizers, $M_{\lambda/4}$ the quarter-wave retarder, $M_1$ to $M_N$ correspond to the retardation matrices containing the information of the previously calculated optical axis along the light path (see fig. \ref{fig:PLI} right).   
   
   \paragraph{Application:}
   As an example a crossing region inspired by the optic chiasm was created (see fig. \ref{fig:sim_example} a)). In this example, two densely packed fiber bundles cross each other (blue curves). In addition, there are two smaller fiber bundles within the large one, each of which changes the bundle in the intersection and thus remains on the respective left or right half of the brain in analogy to the optic chiasm (orange curves). Furthermore, a regular variating radius was created for all fibers, indicating the nodes of Ranvier. Due to the solving of the collision, the crossing region expands. As a result of initial random shifts to the fiber positions, a wave pattern becomes visible.
   
   After the generation of the volume, a middle section is cut out to apply the virtual 3D-PLI measurement. From the simulation, the transmittance, the direction and the retardation maps (see eq. \ref{eq:PLI}) are calculated. From this data it is possible to generate a so called fiber orientation map (FOM) \cite{Axer2011a}. The FOM represents the orientation of the fibers in a coloured space where each color represents a 3D direction (see fig. \ref{fig:sim_example} c) and d)). A comparison of the two simulations clearly shows that collision-free models are necessary.
   
   %------------------------------------------------
   
   \subsection{Towards large scale brain modeling}
   
   The algorithm presented here shows potential for the generation of volumes in the order of $\SI{100x100x100}{\micro\meter}$. This is a sufficient volume size to investigate the influence of the underlying microscopic fiber structure on the resulting 3D-PLI signal. With a mean segment length $\bar{l} = \SI{2}{\mu\meter}$ and a fiber radius of $r = \SI{1}{\mu\meter}$ the computation time on the JURECA-system is around $\SI{10}{\minute}$ with variations depending on the initial complexity of the generated configuration (e.g., large number of fibers or crossings). 
   
   Modeling of large-scale volumes will certainly require modifications of the here presented algorithm. To give some realistic examples, a rat brain section for 3D-PLI measurement spans a volume of about $\SI{1}{\centi\meter} \times \SI{1}{\centi\meter} \times \SI{60}{\micro\meter}$, while a human brain section covers a volume of around $\SI{10}{\centi\meter} \times \SI{10}{\centi\meter} \times \SI{60}{\micro\meter}$. Such large section volumes have to be divided into cubes (i.e., sub-voxels) of the same size and with an edge length of the same length as the section thickness. Consequently, the rat section mentioned above has to be divided into $\sim10^4$ cubes, while the human brain section results in $\sim10^6$ cubes. These cubes can be computed on distributed nodes. However, fibers are likely to traverse several cubes, which requires communication between compute nodes to enable continuous integration of fiber segments from neighboring cubes. 
   
   Further optimization can be achieved by using the more dedicated GPU architecture, such as provided by the JURECA-system. This opens up new possibilities to replace the serially acting octree implementation of the algorithm by a more advanced parallel constructable Z-curved binary tree as described in \cite{Karras}.
   
   %------------------------------------------------
   
   \section{Conclusion}
   
   Simulation has become an indispensable tool in neuroscience, since it provides both overarching (though usually simplified) insights to the observed system and its interactive manipulation. Our aim was to develop a new algorithm to create highly-dense fiber models of nerve fiber bundles for 3D-PLI, which reflect the fiber architecture of the brain.
   
   We presented an algorithm capable of generating complex fiber models containing fiber bundles. These models are created from simple configurations, which are then reconfigured into a collision-free fiber constellation. The transformations that perform these reconfigurations are controlled by boundary parameters. The example of \SI{90}{\degree} crossing fiber bundles clearly demonstrated the strong influence of interpenetrating fibers on the simulated measurements and their fiber orientation-based interpretation. Since all fibers are represented as a list of coordinates and radii, these models can also be used for other simulation approaches, of complementary imaging techniques, such as diffusion MRI.
   
   \subsubsection*{Acknowledgments.}
   The authors gratefully acknowledge the computing time granted by the JARA-HPC Vergabegremium and provided on the JARA-HPC Partition part of the supercomputer \emph{JURECA} at Forschungszentrum J\"ulich.
   
   This project has received funding from the European Union’s Horizon 2020 Research and Innovation Programme under Grant Agreement No. 7202070 (Human Brain Project SGA2).

   %%%%%%%%%%% The bibliography starts:

\end{document}